\documentstyle[amssymb,11pt]{amsart}
\newcounter{sub}
\newcommand{\im}{\mbox{Im}}
\newcommand{\D}{\Delta}

\newcommand{\al}{\alpha}

\newcommand{\CC}{{\cal C}}
\newcommand{\CD}{{\cal D}}
\newcommand{\CF}{{\cal F}}
\newcommand{\CH}{{\cal H}}
\newcommand{\CI}{{\cal I}}
\newcommand{\CK}{{\cal K}}
\newcommand{\cG}{{\cal G}}
\newcommand{\CN}{{\cal N}}  

\newcommand{\CS}{{\cal S}}
\newcommand{\CV}{{\cal V}}
\newcommand{\CW}{{\cal W}}

\newcommand{\CT}{{\cal T}}

\newcommand{\CL}{{\cal L}}

\newcommand{\om}{\Omega}
\newcommand{\la}{\lambda}

\newcommand{\pa}{\partial}
\newcommand{\be}{\begin{equation}}
\newcommand{\eq}{\end{equation}}

\newcommand{\aufz}
{\begin{list}{$\bullet$}{\topsep0cm \itemsep0cm \parsep0cm}}
\newcommand{\eaufz}{\end{list}}
\newcounter{num}
\newcommand{\remlst}{\begin{list}
{(\arabic{num})}{\usecounter{num}\topsep0cm \itemsep0cm \parsep0cm}}
\newcommand{\bh}{{\Bbb H}}

\newcommand{\ho}{\hat{\otimes}}
\newcommand{\gb}{{\goth B}}

\newcommand{\Ga}{\Gamma}

\newcommand{\gn}{{\goth n}}
\newcommand{\GS}{{\goth S}}

\newcommand{\sj}[6]{{\textstyle\!\left\{ {#1\atop #2}{#3\atop #4}
\Big| {#5 \atop #6}\right\}_q}}
\newcommand{\gsj}[6]{{\textstyle\!\left\{\!\!\left\{ {#1\atop #2}{#3\atop #4}
\Big| {#5 \atop #6}\right\}\!\!\right\}_q}}

\newcommand{\sjm}[6]{{\textstyle\!\left\{ {#1\atop #2}{#3\atop #4}
\Big| {#5 \atop #6}\right\}_{q^{-1}} }}
\newcommand{\usl}{{\cal U}_{q}(sl(2))}

\newcommand{\BN}{{\Bbb N}}

\newcommand{\BC}{{\Bbb C}}
\newcommand{\BA}{{\Bbb A}}
\newcommand{\BP}{{\Bbb P}}
\newcommand{\BZ}{{\Bbb Z}}

\newcommand{\rb}[2]{\!{\textstyle \left( {{#1}\atop {#2}} \right) }}
\newcommand{\sbf}[4]{\!{\textstyle \left[ {{#1}\atop {#2}}{{#3}\atop
    {#4}} \right] }}
\newenvironment{proof}{\noindent{\it Proof\/}:}{$\;\Box$}
\newenvironment{definition}{\par\vspace{.5\baselineskip}
\noindent{\bf Definition\/}:}{\par\vspace{.5\baselineskip}}
\newtheorem%
{thm}{Theorem}[section]
\newtheorem%
{proposition}[thm]{Proposition}
\newtheorem%
{lemma}[thm]{Lemma}
\newtheorem%
{lemmadef}[thm]{Lemma-Definition}
\newtheorem%
{corollary}[thm]{Corollary}
\newtheorem%
{conjecture}[thm]{Conjecture}

\begin{document}
\thispagestyle{empty}
\hspace*{\fill} PAR-LPTHE 96-02\\[.5cm]
\begin{center}{ \bf\sc\Large Weak quasitriangular Quasi-Hopf algebra 
structure\\ of minimal models}\\[.5cm]
{\sc J. A. Teschner\footnote{The author thanks the DFG for financial support}}\\[.5cm]
{Laboratoire de Physique Th\'{e}orique et Hautes Energies,\\
Universit\'{e} Pierre et Marie Curie, Paris VI,\\
Universit\'{e} Denis Diderot, Paris VII,\\
Bte 126, 4 Place Jussieu, 75252 Paris cedex 05, France\\
teschner@@lpthe.jussieu.fr}\\[.5cm]
{December 1995}
\end{center}
\begin{abstract}
The chiral vertex operators for the minimal models are constructed and used to define a fusion 
product of representations. The existence of commutativity and associativity operations is proved.
The matrix elements of the associativity operations are shown to be given in terms of the 
6-j symbols of the weak quasitriangular quasi-Hopf algebra obtained by truncating $\usl$
at roots of unity.
\end{abstract}
\section{Introduction}
Structures related to quantum groups encode important information on conformal field theories. 
Whereas the chiral algebra (Virasoro, Kac-Moody, $\CW$ etc.)
may be considered to describe the local properties of the theory, the relation to quantum groups 
nicely describes its global properties such as monodromies of correlation functions
and exchange relations of operators. A nice picture emerges that puts (rational) conformal 
in analogy to classical group theory \cite{MS}\cite{FFK}. 
This analogy may be formulated more precisely in the language of braided tensor categories,
as has been worked out for (negative level) WZNW-models in \cite{KL}. A similiar presentation has 
not yet been 
rigorously worked out in the case of the minimal models. However, various investigations have 
produced a reasonable expectation of what the quantum group relevant for the minimal models should 
be: On one hand side, in \cite{CGR} the operator algebra for Liouville theory at irrational 
central charges was determined. It was shown to be given in terms of the 
6-j symbols of the quantum group $\usl$. However, the truncation
of the operator algebra due to the reducibility of the relevant Virasoro representations was not 
discussed there. In \cite{MaS1} on the other hand side the quantum group structure that is 
compatible with the additional truncation at rational central charge was worked out. It was shown 
to be given by a structure called weak quasitriangular quasi-Hopf algebra, in which 
co-associativity has to be modified to account for the truncation. This structure was shown to be
relevant in the simplest nontrivial example of the Ising model in \cite{MaS2}. \par
In a sense, what remains to be done is to establish the foundations to apply the methods of 
\cite{CGR}\cite{MaS1} to general minimal models. This is one of the aims of the present work.
The main other objective is to introduce a formalism, based on the notion of a fusion product, 
which makes abovementioned analogy to classical group theory (resp. relation to braided tensor
categories) more explicit.\par
One of the technical problems to deal with comes from the fact that there are no explicit 
expressions for the differential equations following from decoupling of general nullvectors.
One therefore has to find indirect ways to obtain the required results.\par
The content of the present paper may be summarised as follows: After a
brief review of results on the relevant Virasoro representations the third section describes the
construction of chiral vertex operators, in particular a rigorous proof of the fusion rules.
Although this has been done already in \cite{FF3}, it may be worthwhile to have an alternative
more elementary approach. A concise description for the Virasoro transformation behaviour of
general descendants is obtained.\par
In the next section the concept of a fusion product of representations is defined from that of
chiral vertex operators. The conformal properties of the fusion product may be expressed in 
a very concise way.\par
The fifth section discusses the composition of fusion products as well as 
definition of correlation functions. The contact to the formalism of \cite{BPZ} is established.\par
Global properties of the fusion product, which are its commutativity and associativity laws, are 
considered in the following section. The main point is to show 
that an associativity law really exists.
Given that, it is not difficult to show that the matrices which represent commutativity and
associativity satisfy the pentagon an hexagon identities of \cite{MS}.
Finally, it is shown how the strategy of \cite{CGR} can be applied to obtain the explicit 
expression for the matrix representing associativity, which is given in terms of the 
truncated q-6j symbols $\usl$ as defined in \cite{KR}.
\section{Verma modules vs. irreducible highest weight modules}
A Verma module $\CV_h$
is defined as the highest weight representation for
which the states
\begin{equation}\label{basis}  L_{-n_1}L_{-n_2}\ldots L_{-n_k}v_h \quad \mbox{with}
\quad k\in\BN
\quad \mbox{and} \quad
n_{i}\ge n_{j} \quad\mbox{for all}\quad i<j\quad i,j=1\ldots k
\end{equation}
form a basis. $\CV_h$ may be decomposed into $L_0$ (energy) eigenstates.
\[ \CV_{h}=\bigoplus_{n\in\BN}\CV_{h}^{(n)}
\qquad\qquad L_0\CV_{h}^{(n)}=(h+n)\CV_{h}^{(n)}.\]
The subspace $\CV_{h}^{(n)}$
with energy $h+n$ is spanned by all vectors of the form
(\ref{basis}) such that $n=\sum_{i=1}^k n_i$. The number $n$ will
be called level in the following.
There is a unique bilinear form $<\! .,.\! >$ on $\CV_h$ such that
$<\!v_h,v_h\!>=1$ and $<\!L_n\xi,\zeta\!>=<\!\xi,L_{-n}\zeta\!>$. One may prove \cite{KaRa}
that its kernel is the maximal proper submodule $\CS_h$ contained in
$\CV_h$. $<\!.,.\!>$ becomes nondegenerate on the {\it irreducible}
representation $\CH_h:=\CV_h/\CS_h$.
Reducibility is equivalent to the existence of vectors ${\goth n}_h$
besides $v_h$ that obey $L_n{\goth n}_h=0$, $n>0$, called {\it null vectors}.
The nullvectors in $\CV_h$ generate the singular submodule $\CS_h$.\par  
The case to be considered will be that of rational central charge
$c=c_{p'p}=1-6\frac{(p'-p)^2}{p'p}$, with $p\in\BN$, $p'\in\BN$ coprime, and representations
with highest weights 
\[ \textstyle
 h\in\GS:=\left\{ h_{j',j}=j'(j'+1)\frac{p}{p'}+j(j+1)\frac{p'}{p}-2j'j-j'-j;
0\leq j'\leq \frac{p'-2}{2}\wedge 0\leq p \leq \frac{p-2}{2}\right\} ,\]
The tuple $(j',j)$ will be abbreviated $J$.
Correspondingly the spaces $\CV_{h_{j',j}}$, $\CH_{h_{j',j}}$, $\CS_{h_{j',j}}$ will be
denoted $\CV_J$, $\CH_J$, $\CS_J$.\par
In this case the structure of Verma-modules may be described as follows \cite{FF1}\cite{FF2}:
The singular submodule $\CS_J$
is generated by two nullvectors in $\CV_J$: One at level $(2j'+1)(2j+1)$,
the other at level $(p'-2j'-1)(p-2j-1)$. However, $\CS_J$ itself is reducible.
One ends up with an infinite nested inclusion of Verma modules, one being
generated by the nullvectors of the other.
The structure of embeddings of Verma modules may be summarized by the following diagram: \\
\unitlength.75cm
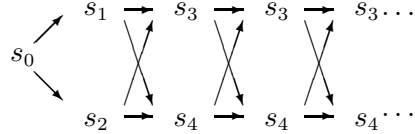
\begin{picture}(10,3)
\put(4,1.5){\parbox{.6cm}{\center $s_0$}}
\put(4.6,1.8){\vector(1,1){.5}}
\put(4.6,1.3){\vector(1,-1){.5}}
\put(5.3,2.3){\parbox[b]{.6cm}{\center $s_1$}}
\put(5.3,0.7){\parbox[t]{.6cm}{\center $s_2$}}
\put(6.2,2.4){\vector(1,0){.5}}
\put(6.2,0.5){\vector(1,0){.5}}
\put(6.2,2.2){\vector(1,-3){.5}}
\put(6.2,0.7){\vector(1,3){.5}}
\put(6.9,2.3){\parbox[b]{.6cm}{\center $s_3$}}
\put(6.9,0.7){\parbox[t]{.6cm}{\center $s_4$}}
\put(7.8,2.4){\vector(1,0){.5}}
\put(7.8,0.5){\vector(1,0){.5}}
\put(7.8,2.2){\vector(1,-3){.5}}
\put(7.8,0.7){\vector(1,3){.5}}
\put(8.5,2.3){\parbox[b]{.6cm}{\center $s_3$}}
\put(8.5,0.7){\parbox[t]{.6cm}{\center $s_4$}}
\put(9.4,2.4){\vector(1,0){.5}}
\put(9.4,0.5){\vector(1,0){.5}}
\put(9.4,2.2){\vector(1,-3){.5}}
\put(9.4,0.7){\vector(1,3){.5}}
\put(10.1,2.3){\parbox[b]{.6cm}{\center $s_3$}}
\put(10.1,0.7){\parbox[t]{.6cm}{\center $s_4$}}
\put(10.7,2.3){\parbox[b]{.6cm}{\center $\ldots$}}
\put(10.7,0.7){\parbox[t]{.6cm}{\center $\ldots$}}
\end{picture}\\
There is an arrow from a vector $s_i$ to another
vector $s_j$ whenever $s_j$ is a generator of the submodule $\CS_{h_i}$ in the Verma module 
generated from the null vector $s_i$. 
\section{Chiral vertex operators}
\subsection{Conditions for existence}
The first aim is to construct operators $\psi_{\bh}(z)$, $\bh={h_2 \choose h_3h_1}$ such that:
\newcommand{\ph}{\psi_{\bh}(z)}
\aufz
\item[(a)] $\ph:\CH_{h_1}\rightarrow\CH_{h_3}$,
\item[(b)] $[L_n,\ph]=z^n(z\pa+h_2(n+1))\ph.$
\eaufz
Start by defining linear forms 
\newcommand{\xid}[1]{{}^t\xi_{\bh}^{(#1)}}
\newcommand{\xiv}[1]{\xi_{\bh}^{(#1)}}   
$\xid{n}$ on $\CV^{(n)}_{h_3}$ by 
\aufz
\item $\xid{0}(v_{h_3})=\CN_{\bh}$, the number $\CN_{\bh}$ being called the normalization of
$\ph$. This is extended to arbitrary $n$ by
\item $\xid{n}(L_{-k}\zeta^{(n-k)})=(\D+n-k+h_2(k+1))\xid{n-k}(\zeta^{(n-k)})$ for any 
$\zeta^{(n-k)}\in\CV_{h_3}^{(n-k)}$, where $\D=h_3-h_2-h_1$.
\eaufz
\begin{proposition}\label{nec}
Necessary conditions for existence of chiral vertex operators $\ph$ are $\xid{n}(\gn)=0$ for
any nullvector $\gn\in\CV_{h_3}^{(n)}$. 
\end{proposition}
\begin{proof} Consider $\ph v_{h_1}$: First observe that 
\begin{equation}
\ph v_{h_1}:= \sum_{n=0}^{\infty}z^{\D+n}\xiv{n}
\end{equation}
is necessary for (b) to hold in the case of n=0. Next it is easy to see that 
\begin{equation}
L_k\xi_{\bh}^{(n)}=(\D+n-k+h_2(k+1))\xi_{\bh}^{(n-k)}\quad\mbox{for}
\quad k\leq n. \label{rec1}\end{equation}
is necessary for (b),$n>0$. One therefore needs to find necessary and sufficient conditions for the
system of equations (\ref{rec1}) to be solvable. It suffices to consider $k=1,2$ since all other 
$L_k$, $k>2$ are generated from these. Suppose a solution exists for any $m<n$. Then consider the
linear map 
\[ A^{(n)}:\quad
\CH_{h_3}^{(n)}\rightarrow \CK_{h_3}^{(n)}\qquad A^{(n)}(\xi^{(n)})=
L_1\xi^{(n)}+L_2\xi^{(n)},\]
where $\CK_{h_3}^{(n)}:= \CH_{h_3}^{(n-1)}\oplus\CH_{h_3}^{(n-2)}$.
If there is no nullvector in $\CV_{h_3}^{(n)}$ then $A^{(n)}$ is invertible and one has a unique 
solution $\xiv{n}$. If there are nullvectors then $\CK^{(n)}_{h_3}=\im A^{(n)}\oplus\CC$ with
a nontrivial $\CC$ and a solution will exist only if the vector 
\newcommand{\lk}{(\D+n-k+h_2(k+1))}
$\sum_{k=1,2}\lk\xiv{n-k}$ has no components in $\CC$. This will be the case iff it is annihilated
by any linear form ${}^t c^{(n)}$ that vanishes on $\im A^{(n)}$. Such linear forms may be 
represented by elements $c^{(n-1)}+c^{(n-2)}\in\CK_{h_3}^{(n)}$ such that 
$<\! c^{(n-1)}+c^{(n-2)},L_1\zeta^{(n)}+L_2\zeta^{(n)}\! >=0$ for any 
$\zeta^{(n)}\in\CH_{h_3}^{(n)}.$
But this is equivalent to the statement that $L_{-1}c^{(n-1)}+L_{-2}c^{(n-2)}$ represents an
element of the singular subspace $\CS_{h_3}^{(n)}$ of $\CV_{h_3}^{(n)}$. The condition for 
existence of solutions now reads
\[ <\! c^{(n-1)}+c^{(n-2)}, \sum_{k=1,2}\lk\xiv{n-k}\!>=0, \]
which is equivalent to the equation obtained by considering the form $<,>$ on the Verma 
module and replacing $c^{(n-1)}+c^{(n-2)}$ by $\gn^{(n-1)}+\gn^{(n-2)}$, where $\gn^{(n-1)}$, 
$\gn^{(n-2)}$ are vectors in $\CV_{h_3}$ such that $\gn=L_{-1}\gn^{(n-1)}+L_{-2}\gn^{(n-2)}$ is
 a null vector in $\CV_{h_3}^{(n)}$. By definition of $\xid{n}$ this is just equivalent to 
$\xid{n}(\gn)=0$.\end{proof} 
\begin{proposition} \label{furu}
{\it Fusion rules:} Let $h_i=h_{J_i}$, $J_i=(j_i',j^{}_i)$, $i=1,2,3$. Then 
a complete set of solutions to the condition that 
$\xid{n}(\gn)=0$ for any nullvector $\gn\in\CV_{h_3}^{(n)}$  
is provided by the triples $(J_1,J_2,J_3)$ such that 
\[\begin{array}{c}
  |j'_2-j'_3|\leq j_1'\leq\min(j'_2+j'_3,p'-2-j'_2-j'_3) \\
  |j_2-j_3|\leq j_1\leq\min(j_2+j_3,p-2-j_2-j_3).
\end{array}\]  
\end{proposition}
\begin{proof} Introduce the notation $PJ=(\frac{p'-2}{2}-j',\frac{p-2}{2}-j)$
The singular subspace $\CD_{h_3}$ of $\CV_{h_3}$ is generated by nullvectors $\gn_{J_3}$, 
$\gn_{PJ_3}$ at level $(2j+1)(2j'+1)$ and $(p-2j-1)(p'-2j'-1)$ respectively. Write them as
\begin{equation}
\gn_J=\sum_{\pi}C_{\pi}L_{-\pi}v_J,\qquad J=J_3\quad\mbox{or}\quad J=PJ_3, 
\end{equation}
where the summation is performed over all vectors $\pi=(n_k,n_{k-1},\ldots,n_1)$ of integers such
that $n_{i+1}\geq n_i$ and $\sum_{k=1}^{k}n_i=(2j'+1)(2j+1)$, whereas 
$L_{-\pi}=L_{-n_k}L_{-n_{k-1}}\ldots L_{-n_1}$. One has 
\[ \xid{n}(L_{-n_k}L_{-n_{k-1}}\ldots L_{n_1}v_J)=
\prod_{i=1}^{k}\left(\D-\sum_{j=1}^{i-1}n_j+h_2(1-n_i)\right). \]
$\xid{n}(\gn_J)$ is of the form 
\begin{equation} \xid{n}(\gn_J)=P_J(h_1,h_2;c),\end{equation}
where $P_J$ is some polynomial in $h_1$, $h_2$. A nice expression for this polynomial 
has been found by Feigin and Fuchs in \cite{FF1}: They consider the following problem: There is a 
simple Virasoro representation on a vector space with basis 
$\{ f_n;\quad n\in\BZ \}$ defined by
$L_{-k}f_n:=(\mu+n-\la(k+1))f_{n+k},\quad cf_n:=0$.What Feigin and
Fuchs calculated is $\gn_J f_0=Q(\mu,\la;J;t)f_{(2j'+1)(2j+1)}$, where $t=-p'/p$.
It is easy to see that $P_J(h_1,h_2;c)$ is proportional to $Q(\D,-h_2;J,t)$. 
The expression given in 
\cite{FF1} may be written as ($\al_{\pm}:=\pm(p'/p)^{\pm 1/2}$):
\[ \big( P_J(h_1,h_2;c) \big)^2 =\!\!\prod_{-j'\leq m'\leq j' \atop
  -j\leq m\leq j}\!\!  
\Big\{ (h_2-h_1)^2+2(h_2+h_1){[}m'\al_-+m\al_+{]}^2+h(m',m)h(-m',-m)\Big\}.
\]
Another useful form of this expression is obtained by introducing
$r_{m'm}=m'\al_-+m\al_+$, and parametrizing
$h_1=\al_1^2-\al_0^2$, $h_2=(\al_1+\delta)^2-\al_0^2$, $2\al_0=\al_++\al_-$: 
\[ P_J(h_1,h_2;c)=\prod_{-j'\leq m'\leq j' \atop
  -j\leq m\leq j}(\delta-r_{m'm})(\delta+2\al_1+r_{m'm})
. \]
The necesary condition for existence of $\ph$ is equivalent to the
vanishing of $P_{J_3}(h_1,h_2;c)$ and $P_{PJ_3}(h_1,h_2;c)$. By writing 
$\al_1=j'_1\al_{-}+j_1\al_+$ one easily reads off
that the set of solutions to $P_{J_3}(h_1,h_2;c)=0$ is given
by \[
\CS_1:=\left\{ h_2=h_{J_2}; J_2=(j'_2,j_2) \left| 
\begin{array}{c}
  {\scriptstyle j'_2=j'_1-j'_3,j'_1-j'_3+1,\ldots,j'_1+j'_3} \\
  {\scriptstyle j _2=j_1 -j_3 ,j_1 -j_3 +1,\ldots,j_1 +j_3 }
\end{array}\right.\right\} \]
The set of solutions to $P_{PJ_3}(h_1,h_2;c)=0$ may similarly be read off by 
parametrizing\footnote{Because of $p'\al_{-}+p\al_{+}=0$ the change 
$j_1\rightarrow \frac{p-2}{2}-j_1$, $j'_1\rightarrow \frac{p'-2}{2}-j'_1$ doesn't
change $\al_1$, but only the labelling of solutions.}
$\al_1=\big( \frac{p'-2}{2}-j'_1\big)\al_{-}+\big( \frac{p-2}{2}-j_1\big)\al_+$.
\[
\CS_1:=\left\{ h_2=h_{J_2}; J_2=(j'_2,j_2) \left| 
\begin{array}{c}
  {\scriptstyle j'_2=j'_3-j'_1,j'_3-j'_1+1,\ldots,p-2-j'_1-j'_3} \\
  {\scriptstyle j _2=j_1 -j_3 ,j_3 -j_1 +1,\ldots,p-2-j_1 -j_3 }
\end{array}\right.\right\} \]
The intersection of these sets may be parametrized as in the Proposition.\end{proof}
\begin{proposition}\label{suff}
The fusion rules are also sufficient for a chiral vertex operator $\ph$ to exist.
\end{proposition}
\begin{proof}
From the proof of Proposition \ref{nec} one finds a unique definition of $\ph v_{h_1}$.
One may recursively extend the definition of $\ph$ to arbitrary $\zeta\in\CV_{h_1}$ by 
defining 
\begin{equation}
\psi_{\bh}(z)
L_{-k}\zeta=-z^{-k}(z\pa_z+h(1-k))\psi_{\bh}(z)\zeta
+L_{-k}\psi_{\bh}(z)\zeta. \label{enexp} 
\end{equation}  
Now condition (b) holds for $n<0$ by definition. The case $n=0$ is trivial, so it remains to 
consider $n>0$. Validitiy of $[L_k,\ph]\zeta_1=z^k(z\pa+h_2(k+1))\ph\zeta_1$ was shown for 
$\zeta_1=v_{h_1}$ in the proof of proposition \ref{nec}. It remains to show that validity for
$\zeta_1=\zeta$ implies validitiy for $\zeta_1=L_{-m}\zeta$. One only needs to use (\ref{enexp})
to to move $L_{-m}$ to the left on both hand sides of 
$[L_k,\ph]L_{-m}\zeta=z^k(z\pa+h_2(k+1))\ph L_{-m}\zeta$ in order to get terms on which the 
assumption may be applied. It is then easy to see that the resulting terms cancel.\par
Up to now the operator $\ph$ was constructed to map from the Verma module $\CV_{h_1}$. It remains 
to be shown that it vanishes on the singular subspace of $\CV_{h_1}$. It is easy to see 
that necessary and sufficient for this is that $<\! v_{h_3},\ph\gn_1\! >=0$ for any nullvector
$\gn_1\in\CV_{h_1}$. As in the proof of Proposition \ref{furu} is is found that
\[ <\! v_{h_3},\ph\gn_J\!>\propto P_J(h_2,h_3;c)\qquad\mbox{where now}\quad J=J_1\quad\mbox{or}
\quad PJ_1. \]
Since the fusion rules are symmetric with respect to permutations of the $J_i$, the evaluation of
\mbox{$<\! v_{h_3},\ph\gn_1\! >=0$} yields no new conditions.\end{proof}
\subsection{Descendant operators}
$\psi_{\bh}(z)$ is just one member of a whole class of operators
$\psi_{\bh}(\xi_{h_2}|z)$ which may be labelled by vectors $\xi\in
\CV_{h_2}$.
They are recursively defined by 
\begin{equation} \label{desdef}\begin{array}{c}
\psi_{\bh}(v_{h_2}|z):=\psi_{\bh}(z) \qquad\qquad
\psi_{\bh}(L_{-1}\xi|z):=\pa\psi_{\bh}(\xi|z) \\
\psi_{\bh}(L_{-n}\xi|z):=\frac{1}{(n-2)!}
(\pa^{n-2}T_{<}(z)\psi_{\bh}(\xi|z)+\psi_{\bh}(\xi|z)
 \pa^{n-2}T_{>}(z))
\end{array} \end{equation}
where $n\geq2$, and 
\begin{equation} T_{<}(z):= \sum_{n=0}^{\infty}z^n L_{-n-2} \qquad\qquad
T_{>}(z):=\sum_{n=1}^{\infty}z^{-n} L_{n-2}. \end{equation} 
Instead of considering $\psi_{\bh}(\xi|z)$ as an operator that maps
from $\CH_{h_1}$ to $\CH_{h_3}$, one may fix an element
$\zeta\in\CH_{h_1}$ and consider $\psi_{\bh}(\;.\;|z)\zeta$ as an
operator that maps from $\CV_{h_2}$ to $\CH_{h_3}$. In the next subsection,
a formalism will be presented that puts both points of view on equal
footing.\par   
The following theorem gives a convenient characterization of the
conformal properties of the descendants (I will omit the subscript
$\bh$ in the following):
\begin{thm}\label{dtrf}
\[ [L_n,\psi(\xi|z)]
=\sum_{k=-1}^{l(n)}z^{n-k}\left(
{n+1 \atop k+1}\right) \psi(L_k\xi_h|z) \quad \mbox{\it where}
\quad l(n)=\left\{ \begin{array}{r@{\quad\mbox{\it for}\quad}l}
n & n\geq -1 \\ \infty & n<-1. \end{array} \right.
\]\end{thm}
Before giving the proof, I want to explain its implications:
The content of the theorem may be summarized even more concise in
the following rules:
\begin{eqnarray}
{[}T_>(u),\psi(\xi|z){]} &=& \psi(T_>(u-z)\xi|z) \label{tmov1} \\
{[}\psi(\xi|z),T_<(u){]} &=& \psi(T_>(u-z)\xi|z) \label{tmov2} \\
\psi(T_<(u-z)\xi|z) &=& T_<(u)\psi(\xi|z)+\psi(\xi|z)T_>(u), \label{tmov3}
\end{eqnarray}
These formulae allow to write down the complete operator product
expansion of $T(u)\psi(\xi|z)$:
\begin{eqnarray}
T(u)\psi(\xi|z) & = & [T_>(u),\psi(\xi|z)]+T_<(u)\psi(\xi|z)
+\psi(\xi|z)T_>(u) \nonumber \\
 & = & \sum_{k=-\infty}^{\infty}(u-z)^{-k-2}\psi(L_k\xi|z) \\
 & = & \psi(T(u-z)\xi|z). \nonumber
\end{eqnarray}
The sum is finite if $\xi$ contains only finitely many $L_{-n}$
generators. It is now possible to make contact with the more usual
formulations of conformal field theories \cite{BPZ}: One has
\begin{eqnarray}
\psi(L_{-n}\xi|z) & = & \mbox{Res}_{u-z}[(u-z)^{-n+1}T(u)\psi(\xi|z)] \label{descres}\\
 & = & \oint\frac{du}{2\pi i}(u-z)^{-n+1}T(u)\psi(\xi|z)\label{descint}.
\end{eqnarray}
In \cite{BPZ}, these equations are used to {\it define} the
formalism. In the present formalism they have been {\it derived}
purely algebraically.\\
\begin{proof}
As a preliminary note that an alternative basis for $\CV_{h_2}$ may
be written as follows: Let $\vec{n}=(n_1,\ldots ,n_k)$ be a vector of
integers with $n_1\geq 0$, $n_k\geq 0$ and $n_i>0$ for $i=2\ldots k-1$.
Then a basis for $\CV_{h_2}$ is given by the set of all
\begin{equation}
L_{-1}^{n_1}L_{-2}^{n_2}L_{-1}^{n_3}\ldots L_{-1}^{n_{k-1}}
L_{-2}^{n_k}v_{h_2}    \label{base}
\end{equation}
It therefore suffices to use (\ref{desdef}) for $n=1,2$ only.
The theorem will be proved for vectors $\xi$ of the form (\ref{base})
by induction on the integer $s$,
defined as $s:=\sum_{i=1}^k n_i$.
For $s=0$ one easily recognizes the theorem as the covariant
transformation law of $\psi(z)$. Now assume that the theorem holds for
$\psi(\xi|z)$. Consider first $[L_n,\psi(L_{-1}\xi|z)]$: By using the
definition of $\psi(L_{-1}\xi|z)$ and the inductive assumption this is calculated as:
\[
{[}L_n,\psi(L_{-1}\xi|z){]}=
\sum_{k=-1}^{l(n)}z^{n-k}\left( {n+1 \atop k+1}\right)\left(
\frac{n-k}{z} \psi(L_k\xi|z)+\psi(L_{-1}L_k\xi|z) \right)
\]
The first sum may be rewritten as
$\sum_{k=0}^{l(n)}z^{n-k}\left( {n+1 \atop k+1} \right)
\psi([L_k,L_{-1}]\xi|z)$ which proves the theorem for $\psi(L_{-1}\xi|z)$.\par
The key step for the computation of $[L_n,\psi(L_{-2}\xi|z)]$ 
is contained in the following:
\begin{lemma}
\[ 
{[}L_n,T_<(z){]}\psi(\xi|z)+\psi(\xi|z){[}L_n,T_>(z){]}
=\sum_{k=-1}^{l(n)} z^{n-k}\left({n+1 \atop k+1}\right) \psi({[}L_k,L_{-2}{]}\xi|z)
\]
\end{lemma}
\begin{proof}
One has to distinguish
two cases: $n\geq -1$ and $n<-1$. In the first case use
\[ \begin{array}{l}\displaystyle
{[}L_n,T_{<}(z){]}=z^n(z\pa +2(n+1))T_{<}(z)+\sum_{m=1}^{n}z^{n-m}(2n-m+2)
L_{m-2} + \frac{c}{12}n(n^2-1)z^{n-2} \\
\displaystyle
{[}L_n,T_{>}(z){]}=z^n(z\pa +2(n+1))T_{>}(z)-\sum_{m=1}^{n}z^{n-m}(2n-m+2)
L_{m-2},
\end{array}\]
with the convention that $\sum_{m=1}^n(\ldots )=0$ if $n<1$ to find
\[ \begin{array}{cr}\multicolumn{2}{c}{
{[}L_n,T_<(z){]}\psi(\xi|z)+\psi(\xi|z){[}L_n,T_>(z){]} = 
z^{n+1}\psi(L_{-3}\xi|z)+2(n+1)z^n\psi(L_{-2}\xi|z)+ }\\[1ex]
& \displaystyle
+\frac{c}{12}n(n^2-1)z^{n-2}+\sum_{m=-1}^n z^{n-k}(2n-m+2)[L_{m-2},\psi(\xi|z)], 
\end{array} \]
where the sum on the right hand side is evaluated as
\begin{eqnarray*}
 S & := &\sum_{m=1}^n z^{n-m}(2n-m+2)\sum_{k=-1}^{m-2}z^{m-2-k}
\left( {m-1 \atop k+1} \right) \psi(L_k\xi|z) \\
  & = & \sum_{k=1}^n z^{n-k} \left\{ \sum_{m=k}^n(2n+2-m)\left(
        {m-1 \atop k-1} \right) \right\} \psi(L_{k-2}\xi|z).
\end{eqnarray*}
One may prove by induction that the sum within the curly brackets
equals $(k+2)\left( {n+1 \atop k+1} \right)$, so that
\be S=\sum_{k=1}^n z^{n-k}\left({n+1 \atop k+1}\right) \psi(
[L_k,L_{-2}]\xi|z)-\frac{c}{12}n(n^2-1)z^{n-2} \end{equation}
Collecting the different terms one finds the claimed result.The case $n<1$ proceeds analogously.
\end{proof}\\[1ex]
Given the lemma one only needs to apply the inductive assumption for the computation of
$T_<(z){[}L_n,\psi(\xi|z){]}+{[}L_n,\psi(\xi|z){]}T_>(z)$.
\end{proof}
\subsection{Operator differential equations} 
\begin{proposition} The operators $\ph$ satisfy differential equations obtained by 
using (\ref{desdef}) to evaluate in terms of $\ph$ the condition
$\psi_{\bh}(\gn|z)=0$ for $\gn$ being any of the two
nullvectors $\gn_{J_2}$, $\gn_{PJ_2}$ that generate the singular subspace of
$\CV_{h_2}$.
\end{proposition}
\begin{proof} By using (\ref{desdef})one may find
that $<\! v_{h_3},\psi_{\bh}(\gn|z)v_{h_1}\! >$ is proportional to $P_J(h_1,h_3;c)$,
$J=J_2$ or $J=PJ_2$, which vanishes whenever the fusion rules are satisfied.
Furthermore, it is an immediate consequence of Theorem \ref{dtrf} that 
\[ {[}L_n,\psi_{\bh}(\gn|z){]}=z^n(z\pa+(h_2+l)(n+1))\psi_{\bh}(\gn|z),\]
where $l$ denotes the level of $\gn$. But this means that any matrix element 
of $\psi_{\bh}(\gn|z)$ may be expressed as some differential operator acting on 
$<\!v_{h_3},\psi_{\bh}(\gn|z)v_{h_1}\! >$. Since $P_J(h_1,h_3;c)=0$, the operator 
vanishes alltogether.\end{proof}\\[1ex]
In the simplest nontrivial case of $J_2=(0,1/2)$ one gets  
\begin{equation} \pa^2\psi_{\bh}(z)=\al_+^2(T_<(z)\psi_{\bh}+
\psi_{\bh}(z)T_>(z)), \label{opde} \end{equation}
which provides the link of the present treatment to Liouville theory as treated in 
\cite{GN}\cite{CGR}.
\section{Fusion product}
\subsection{}
Consider $\psi_{\bh}(\xi_{2}|z)\xi_{1}$, $\xi_i\in\CH_{h_i}$, $i=1,2$:
Instead of viewing it as the action
of an operator on some state one may view it as the result of taking
some kind of product of two states:
\begin{equation} \psi_{\bh}(\xi_{2}|z)\xi_{1}:=
    {[}\xi_2(z)\ho\xi_{1}(0){]}_{h_3} \end{equation}
The state $\xi_2$ is considered to be located at $z$, $\xi_{1}$ at
$0$. In order to make this more precise I will now consider the action of the translation 
operator $e^{zL_{-1}}$ on states:
\begin{equation}
\xi(z):=e^{zL_{-1}}\xi.\end{equation}
In fact, translated states are nothing new:
\begin{lemma}
For $\bh=\rb{h}{h\; 0}$ one has $\xi(z)=\psi_{\bh}(\xi|z)v_0$.
\end{lemma}
{\it Proof:} This may be verified by noting that\aufz
\item one has $v_h(z)=\psi_{\bh}(z)v_0$ since $v_h(z)$
  satisfies $L_kv_h(z)=z^k(z\pa+h(k+1))v_h(z)$, $k\geq 0$, which are
  the conditions used to define $\psi_{\bh}(z)v_0$ in the proof of Proposition \ref{nec},
  and that
\item  $(L_{-1}\xi)(z)=\pa\xi(z)$, $(L_{-2}\xi)(z)=T_{<}(z)\xi(z)$,
  from which the Lemma may be inductively proved for
  arbitrary $\xi$.\hfill\vspace{1ex}$\Box$
\end{list}
The conformal properties of translated states may be conveniently
summarized by 
\begin{equation} T_{>}(u)\xi(z)=(T_>(u-z)\xi)(z) \qquad\quad
(T_<(u-z)\xi)(z)=T_<(u)\xi(z). \label{conftrst}\end{equation}
Let me also mention the following important special case of (\ref{conftrst}):
\begin{equation} T_>(u)v_h(z)=\left(\frac{h}{(u-z)^2}+
\frac{1}{u-z}\pa\right) v_h(z). \label{tm5} \end{equation}
\begin{definition} The fusion products $[.\ho.]_{h_{3}}$
of two representations $\CH_{h_1}$ and $\CH_{h_2}$ are defined as 
$z_1$, $z_2$ dependent bilinear maps
\[ \begin{array}{rccl}
 [.\ho.]_{h_{3}}: & \CH_{h_1}\otimes\CH_{h_2} & \longrightarrow & \CH_{h_{3}} \\ 
 & \xi_1\otimes\xi_2 & \longrightarrow & [\xi_{2}(z_2)\hat{\otimes}\xi_{1}(z_1)]_{h_3}:=
\psi_{\bh}(\xi_{2}|z_2)\xi_{1}(z_1).\end{array}\] 
\end{definition}
The concept of the fusion product is completely equivalent to that of
the chiral vertex operator: Having defined a fusion product, one may for any fixed $\xi_2$ define 
\[ \begin{array}{rccl} 
 \psi_{\bh}(\xi_{2}|z_2): & \CH_{h_1} & \longrightarrow & \CH_{h_3} \\
 & \xi_1 & \longrightarrow & \psi_{\bh}(\xi_{2}|z_2)\xi_1:= 
[\xi_{2}(z_2)\hat{\otimes}\xi_{1}(0)]_{h_3}. \end{array}\]
\subsection{Local properties of the fusion product}
The conformal properties of $\psi(\xi|z)$ derived above may now be
rephrased as rules for moving $T(u)$ within fusion products:
\begin{eqnarray}
{[}\xi(z)\ho T_<(u)\zeta(z'){]}&=&T_<(u){[}\xi(z)\ho\zeta(z'){]}
+{[}T_>(u)\xi(z)\ho\zeta(z'){]} \label{tm1} \\
{[}T_<(u)\xi(z)\ho\zeta(z'){]}&=&T_<(u){[}\xi(z)\ho\zeta(z'){]}
+{[}\xi(z)\ho T_>(u)\zeta(z'){]} \label{tm2} \\
T_>(u){[}\xi(z)\ho\zeta(z'){]}&=&{[}T_>(u)\xi(z)\ho\zeta(z'){]}
+{[}\xi(z)\ho T_>(u)\zeta(z'){]} \label{tm3},
\end{eqnarray}
which have to be supplemented by (\ref{conftrst}). Since $T_{>}(u)\xi(z)$ will always involve a
derivative with respect to $z$, one may view these relations as describing the response of the 
fusion product with respect to infinitesimal changes of the parameters $z,z'$, i.e. the local 
properties of the fusion product.
\section{Composition of fusion products; Conformal blocks}
\subsection{Triple products}
On the level of formal power series it is possible to define
repeated products such as
\begin{equation}
[[\xi_{3}(z_3)\ho\xi_{2}(z_2)]_{h_{32}}\ho\xi_{1}(z_1)]_{h}
\quad\mbox{or}\quad
[\xi_{3}(z_3)\ho[\xi_{2}(z_2)\ho\xi_{1}(z_1)]_{h_{21}}]_{h},
\label{ex1} \end{equation} 
where $\xi_i\in\CH_{h_i}$, $i=1,2,3$. Consider
i.e. $[[\xi_{3}(z_3)\ho\xi_{2}(z_2)]_{h_{32}}\ho\xi_{1}(z_1)]_{h}$,
where $\xi_i$ will be assumed to have definite level $n_i$:
By definition, the inner bracket $[\xi_{3}(z_3)\ho\xi_{2}(z_2)]_{h_{32}}$ 
may be written as the formal series 
\[ {[}\xi_{3}(z_3)\ho\xi_{2}(z_2){]}_{h_{32}}=\sum_{n=0}^{\infty}
(z_3-z_2)^{\D_{32}+n}\xi_{32}^{(n)}(z_2), \] 
where $\D_{32}=h_{32}-h_2-n_2-h_3-n_3$ and
$\xi_{32}^{(n)}\in\CH_{h_{32}}^{(n)}$. Then the iterated product is defined as 
\begin{eqnarray}
\lefteqn{
  [[\xi_{3}(z_3)\ho\xi_{2}(z_2)]_{h_{32}}\ho\xi_{1}(z_1)]_{h}:=} \nonumber \\
& = & \sum_{n=0}^{\infty} (z_3-z_2)^{\D_{32}+n} 
[\xi_{32}^{(n)}(z_2)\ho\xi_1(z_1)]_h \nonumber \\
& = & \sum_{n=0}^{\infty} (z_3-z_2)^{\D_{32}+n} 
\sum_{m=0}^{\infty} (z_2-z_1)^{\D_{3(21)}+m-n} 
\xi^{(n,m)}_{(32)1}(z_1) \nonumber \\
& = & (z_3-z_2)^{\D_{32}}(z_2-z_1)^{\D_{(32)1}}
\sum_{n=0}^{\infty}\left(\frac{z_3-z_2}{z_2-z_1}\right)^n
\sum_{m=0}^{\infty}(z_2-z_1)^m\xi^{(n,m)}_{(32)1}(z_1)
\label{triple}\end{eqnarray}
where $\D_{(32)1}=h-h_{32}-h_1-n_1$ and $\xi^{(n,m)}\in\CH^{(m)}_h$. 
The sum over $m$ is a formal sum over the homogeneous components $\CH^{(m)}_h$. It may therefore be
exchanged with the summation over $n$ which then defines a vector in 
$\CH^{(m)}_h$ if it converges:
\[ \xi^{(m)}_{(32)1}(z_3,z_2,z_1):= (z_3-z_2)^{\D_{32}}
\sum_{n=0}^{\infty}\left(\frac{z_3-z_2}{z_2-z_1}\right)^n\xi^{(n,m)}(z_1)\]
It may be useful to see what the different possible ways of iterating the fusion 
product correspond to in the language of chiral vertex operators:
\begin{eqnarray} 
{[}\xi_3(z_3)\ho{[}\xi_2(z_2)\ho\xi_1(z_1){]}_{h_{21}}{]}_{h} 
& = & \psi\rb{h_3}{h\;h_{21}}(\xi_3|z_3)\psi\rb{h_2}{h_{21}\;h_1}(\xi_2|z_2)
\xi_1(z_1) \label{A(BC)}\\
{[}{[} \xi_3(z_3)\ho
\xi_2(z_2){]}_{h_{32}}\!\ho\xi_1(z_1){]}_{h} 
&=& \sum_n(z_3-z_2)^{\D_{23}+n}
\psi\rb{h_{32}}{h\;h_1}\!(\xi_{23}^{(n)}|z_2)\xi_1(z_1).\label{(AB)C}
\end{eqnarray}
The order $[A[BC]]$ of taking the fusion product therefore simply
corresponds to the composition of chiral vertex operators, whereas the
expression on the 
r.h.s of the second line has the form one expects the terms of the
operator product expansion of 
$\psi\rb{h_3}{h\;h_{21}}(\xi_3|z_3)\psi\rb{h_2}{h_{21}\;h_1}(\xi_2|z_2)$
to have \cite{MS}.
\subsection{Correlation functions}
Since any multiple product will have to be understood as a formal series over the 
homogeneous components $\CH_{h}^{(m)}$ of some $\CH_{h}$, the definition of higher iterated
products is equivalent to the definition of their matrix elements with arbitrary 
$\zeta^{(m)}\in\CH_{h}^{(m)}$. In the case of triple products one has the four point functions 
such as 
\[ <\! \zeta^{(m)},{[}{[}\xi_3(z_3)\ho\xi_2(z_2){]}_{J_{32}}\ho\xi_1(z_1){]}_{J_{4}}\! >
=(z_2-z_1)^{\D_{(32)1}+m}<\! \zeta^{(m)},\xi_{3(21)}^{(m)}(z_3,z_2,z_1)\! > \]
In the five point case such as 
$<\! \xi_5,{[}\xi_4(z_4)\ho {[}\xi_3(z_3)\ho {[} \xi_2(z_2)\ho \xi_1(z_1)
{]}_{J_{21}}{]}_{J_{3(21)}}{]}_{J_5}\! >$
one finds an apparent ambiguity: it can be expressed as a sum over four point functions
in two ways: either 
\[ \begin{array}{l}\displaystyle
\sum_{n=0}^{\infty}(z_2-z_1)^{\D_{21}+n}
<\! \xi_5,{[}\xi_4(z_4)\ho{[}\xi_3(z_3)\ho\xi^{(n)}_{21}(z_1){]}_{J_{3(21)}}{]}_{J_{5}}\! >
\quad\mbox{or}\\   \displaystyle
\sum_{m=0}^{\infty}(z_3-z_1)^{\D_{3(21)}+n}
<\! \xi_5,{[}\xi_4(z_4)\ho \xi_{3(21)}^{(m)}(z_3,z_2,z_1){]}_{J_5}\!> 
\end{array} \]
In the first case one ends up with a power series of the form
\[ z_{21}^{\D_{21}}z_{31}^{\D_{31}}z_{41}^{\D_{4(3(21))}}\sum_{n=0}^{\infty}\left(
\frac{z_{21}}{z_{31}}\right)^n\sum_{m=0}^{\infty}\left(\frac{z_{31}}{z_{41}}\right)^n
<\! \xi_5,{[}\xi_4(z_4)\ho\xi_{3(21)}^{(n,m)}(z_1)]_{J_5}\!>,\]
where $z_{ij}:= z_i-z_j$, whereas in the second case one has a similiar power series with
exchanged summations over $n$ and $m$. There is no ambiguity if the power series converge to
holomorphic functions, since then the summations can be freely interchanged. On the level of 
formal power series one may remove the ambiguity by adopting the convention that the series
expansion is performed starting from the innermost brackets to the outermost ones.
\subsection{Conformal Ward identities; Decoupling equations}
It is now easy to make contact to the formulation of \cite{BPZ}:
Consider correlation functions such as
\begin{equation}\label{multi} <\!v_0,[\ldots[\xi_1(z_1)\hat{\otimes}\ldots
[\xi_{i}(z_{i})\hat{\otimes}\xi_{i+1}(z_{i+1})]_{h^1}\ldots]\ldots ]_0\!>
\end{equation}
It is useful to have a more concise notation:
The possible multiplications of states $\xi_{i}$, $i=1\ldots n$.
are characterized by the following data:
\begin{enumerate} 
\item A permutation
$\sigma(i)$, $i=1\ldots n$ of $(1,\ldots ,n)$, 
\item a complete binary bracketing of $\sigma(1)\ldots\sigma(n)$ such as
$(((3,5),((1,4),6)),2)$,  
\item The set of tuples $(h_i,\xi_i,z_i)$, where $\xi_i$
is a state in the Verma module of conformal weight $h_i$ and
$z_i$ is the position where the state is supposed to be inserted,
and finally 
\item a set of real numbers $h^r$, $r=1,\ldots , n-1$
associated with each pair of brackets which denote the weights of
the "intermediate" representations appearing in the multiplication.
\end{enumerate}
Let $\gb_n$ denote the set of all collections of data (1)-(2), i.e.
of all bracketings $((1,4),\ldots$. The elements of $\gb_n$ will
be denoted by $\CT,\CT'$ etc..
The tuples $(h^1,\ldots,h^{n-1}:= 0)$
will be abbreviated as $H$.
One may distinguish the 'external' data $(h_i,\xi_i,z_i)$ from the
'internal' data $\Ga:=(\CT,H)$, which parametrize the possible ways
to form fusion products of $\xi_i(z_i)$. \par
Correlation functions such as (\ref{multi}) may then be abbreviated as 
$ <\! \xi_1(z_1)\ldots \xi_n(z_n)\!>_{\Ga}$.\par 
In order to derive the conformal Ward identities consider 
\[ <\! \xi_1(z_1)\ldots (L_{-n}\xi_i)(z_i)\ldots \xi_n(z_n)\!>_{\Ga}. \]
By using (\ref{conftrst}) one  rewrites $(L_{-n}\xi_i)(z_i)$ as 
$\oint\frac{du}{2\pi i}(u-z_i)^{-n+1}T_{<}(u)\xi_i(z_i)$, 
where the integral is to be understood as the operation of taking the residue
in the formal power series. By using (\ref{tm1})-(\ref{tm3}) one moves $T_{<}(u)$  until one gets
\[ \sum_{j\neq i}\oint\frac{du}{2\pi i}(u-z_i)^{-n+1}<\! \xi_1(z_1)\ldots 
(T_{>}(u)\xi_j(z_j))\ldots \xi_n(z_n)\!>_{\Ga}. \]
Using (\ref{conftrst}) again one gets terms with $\oint\frac{du}{2\pi i}(u-z_i)^{-n+1}
(T_{>}(u-z_j)\xi_j)(z_j)$. Expanding $T_{>}(u-z_j)$ one finds precisely what the conformal 
Ward identities of \cite{BPZ} amount to. Besides derivatives with respect to $z_j$ this will only
produce generators $L_n$, $n\geq 0$ which preserve or lower the degree of the homogeneous
components of $\xi_j$. By repeated application of these operations one recovers the fact that
any correlations function $ <\! \xi_1(z_1)\ldots \xi_n(z_n)\!>_{\Ga}$ may be expressed in terms
of meromorphic differential operators acting on $ <\! v_1(z_1)\ldots v_n(z_n)\!>_{\Ga}$,
which are usually called the conformal blocks.\par
One now also immediately gets the differential equations that the conformal blocks have to
satisfy:
First of all, from
$0=<\! L_{k}v_0,[\ldots]_0\!>_{\Ga}$ for $k=-1,0,1$ one finds the equations expressing projective
invariance of the conformal blocks:
\begin{equation} \label{proj}\sum_{i=1}^n z_i^k\left(z_i\frac{\pa}{\pa z_i}+(k+1)h_i\right)
<\! v_1(z_1)\ldots v_n(z_n)\!>_{\Ga} =0;\qquad k\in\{ -1,0,1 \}. \end{equation}
In addition one gets partial differential equations from decoupling of the null vectors in
$\CV_{J_i}$, $i=1,\ldots,n$. If the nullvector in $\CV_{J_i}$ is written as
in the proof of Proposition \ref{furu}, then the resulting differential equation will be
\begin{equation}\label{deco} \begin{array}{c}\displaystyle 
\sum_{\pi}C_{\pi}\CL^{(i)}_{-\pi}<\! v_1(z_1)\ldots v_n(z_n)\!>_{\Ga}=0, \\[1ex] 
\mbox{where}\quad\CL^{(i)}_{-\pi}:= \CL^{(i)}_{-n_k}\ldots\CL^{(i)}_{-n_1},
\displaystyle
\quad\CL_{-n}^{(i)}:= \sum_{j\neq i}\left( \frac{h_{j}(1-n)}{(z_i-z_j)^m}+
\frac{1}{(z_i-z_j)^{m-1}}\pa_j \right) \end{array} \end{equation}
The conformal blocks as constructed from fusion products provide {\it formal power series 
solutions} to these differential equations.
By using standard results on partial differential equations with regular singular points such as
given in \cite{Kn}, Appendix B,
it should be easy to prove that the formal power series in question actually
converge and may be analytically continued to multivalued analytic functions on the
complement of the hyperplanes $z_i=z_j$ in $\BC^n$.   
\section{Global properties of the fusion product: Commutativity and associativity}
\subsection{Commutativity}
The logarithm used to define $(z_2-z_1)^{h_{21}-h_1-h_2}$.
is taken to be the principal value. One therefore
has to distinguish two zones: 
$\BC_+^2:=\{(z_2,z_1)\in\BC^2|\arg(z_2-z_1)\in(0,\pi{]}\}$ and 
$\BC_-^2:=\{(z_2,z_1)\in\BC^2|\arg(z_2-z_1)\in(-\pi,0{]}\}$.
\begin{equation} {[}\xi_2(z_2)\ho\xi_1(z_1){]}_{h_{21}}=\left\{
\begin{array}{c@{\quad\mbox{for}\quad (z_2,z_1)\quad\mbox{in}\quad}c}
e^{ i\pi(h_{21}-h_1-h_2)}{[}\xi_1(z_1)\ho\xi_2(z_2){]}_{h_{21}}&\BC_+^2 \\
e^{-i\pi(h_{21}-h_1-h_2)}{[}\xi_1(z_1)\ho\xi_2(z_2){]}_{h_{21}}&\BC_-^2.
\end{array} \right\}\end{equation}
The phase factor will in the following be abbreviated by 
\begin{equation} \om\rb{h_{21}}{h_2\; h_1}=e^{\pi i(h_{21}-h_1-h_2)}. \end{equation}
\subsection{Associativity}
\begin{thm} \label{asso}
To any four conformal dimensions $h_i=h_{J_i}$ there exists an
invertible matrix $F$ with elements
$F_{J_{21}J_{32}}\sbf{J_3}{J_4}{J_2}{J_1}$ such that
\[
{[}\xi_3(z_3)\ho{[}\xi_2(z_2)\ho\xi_1(z_1){]}_{J_{21}}{]}_{J_4} =
\sum_{J_{23}}F_{J_{21}J_{32}}\sbf{J_3}{J_4}{J_2}{J_1} 
{[}{[} \xi_3(z_3)\ho
\xi_2(z_2){]}_{J_{32}}\!\ho\xi_1(z_1){]}_{J_4},
\] 
for any $\xi_i\in\CH_{h_i}$, $i=1,2,3$. The summation ranges over all $J_{23}$ such that 
the triples $(J_{23},J_2,J_3)$ and $(J_4,J_1,J_{23})$ satisfy the fusion rules.
\end{thm}
\begin{proof} The
Theorem will be deduced (Proposition \ref{reduce}) 
from the corresponding statement for the four-point
conformal blocks 
\begin{eqnarray}
\cG^{J_{21}}_{3(21)}(z_1,z_2,z_3,z_4)& := &
<\! v_0,{[}v_4(z_4)\ho {[}v_3(z_3)\ho{[}v_2(z_2)\ho
v_1(z_1){]}_{J_{21}}{]}_{J_4} {]}_0\! > 
\\
\cG^{J_{32}}_{(32)1}(z_1,z_2,z_3,z_4)&:=&
<\! v_0,{[} v_4(z_4) \ho {[}{[} v_3(z_3)\ho v_2(z_2){]}_{J_{32}}\!\ho
v_1(z_1){]}_{J_4}{]}_0\! >.
\end{eqnarray}
As a preliminary for the proof one needs results on the differential equations that 
$\cG^{J_{21}}_{3(21)}$ and $\cG^{J_{32}}_{(32)1}$ satisfy. 
The three equations (\ref{proj}) determine $\cG$ to be of the form 
\begin{equation} \cG(z_1,z_2,z_3,z_4)=\prod_{i>j}(z_i-z_j)^{\bar{h}-h_i-h_j} 
F\left(\frac{(z_2-z_1)(z_4-z_3)}{(z_3-z_1)(z_4-z_2)} \right),
\label{4ptproj}  \end{equation} where $\bar{h}=\frac{1}{3}\sum_{i=1}^{4}h_i$.
In addition one has eight partial differential equations from null vector decoupling, cf.
(\ref{deco}). The crucial result is then the following:
\begin{proposition}\label{completeness}
$\{\cG^{J_{21}}_{3(21)}\}$ and $\{\cG^{J_{32}}_{(32)1}\}$ form complete sets of solutions of the 
11 ordinary differential equations that follow from null vector decoupling and projective
invariance.
\end{proposition}
\begin{proof} As preparation I will need two Lemmas that are proved in the appendix.
\begin{lemma}\label{fuchsian}
The ordinary differential equation that follows from the decoupling of a nullvector 
$\gn_i\in\CV_{J_i}$, $i=1,2,3,4$  by using (\ref{4ptproj}) are of fuchsian type
(regular singular points at $z_i=z_j, i\neq j\in\{1,2,3,4\}$ and $z_i=\infty$ only, 
see i.e. \cite{CL}), i.e. of the form 
\[ \sum_{k=0}^{l}q^{(k)}(z_i|z_j;j\neq i)\pa^k_i \cG=0\quad\mbox{where}\quad
q^{(k)}=\prod_{j\neq i}(z_i-z_j)^{-(n-k)}p^{(k)}(z_i|z_j;j\neq i),\]
where $l$ is the level at which the nullvector occurs and
$p^{(k)}(z_i|z_j;j\neq i)$ is a polynomial in $z_i$ of order $2(n-k)$ 
(in particular $q^{(n)}=$const.). 
\end{lemma}
\begin{lemma}\label{indicial}
The indicial equation of the fuchsian differential equation from a nullvector in $\CV_{J_i}$ at 
the singular point $z_j$ is equal to $P_{J_i}(h_{J_{ij}},h_{J_{j}};c)=0$.
\end{lemma}
Consider the two differential equations from the nullvectors in $\CV_{J_1}$,
which will be denoted $(J_1)$ and $(PJ_1)$. Let $S_1$ ($S_2$) be the set of roots of  
indicial equation for $(J_1)$ (resp. $(PJ_1)$), and denote $\mu_1(s)$, ($\mu_2(s)$) the multiplicity 
of the root $s\in S_1$ ($s\in S_2$). By the Frobenius method in the theory of ordinary differential 
equations of fuchsian type, see \cite{CL}, it is shown that there is a basis 
$\{ f_{s,k}; s\in S_1,k=0,\ldots,\mu_1(s)\}$ of solutions to equation 
$(J_1)$ with leading asymptotics $(z_2-z_1)^{s}\log(z_2-z_1)^k$. The number
$s$ is called exponent of the solution $f_{s,k}$.
The solutions $f_{s,k}$ are of the 
form $(z_2-z_1)^{s}\log(z_2-z_1)^k g_{s,k}(z_2-z_1)$, where $g_{s,k}(z)$ may be expanded as 
\[ g_{s,k}(z)=1+\sum_{i=1}^{\infty}\sum_{j=0}^{M}a_{ij} z^i(\log(z))^j \]
with some $M\in\BZ^{\geq 0}$. The logarithmic terms may appear (but must not) only if for given $s$ there
is a $s'\in S_1$ such that $s-s'$ is a positive integer.\par
In the present case the roots of the indicial equations for $(J_1)$ and $(PJ_1)$ are parametrized by
the index sets 
\[ \begin{array}{l}
\CI_1:=\left\{ J=(j',j)| 
\begin{array}{c}
  {\scriptstyle j'=j'_2-j'_1,j'_2-j'_1+1,\ldots,j'_1+j'_2-1} \\
  {\scriptstyle j =j_2 -j_1 ,j_2 -j_1 +1,\ldots,j_1 +j_2 -1}
\end{array}\right\}  \\[1ex] 
\CI_2:=\left\{ J=(j',j)| 
\begin{array}{c}
  {\scriptstyle j'=j'_1-j'_2,j'_1-j'_2+1,\ldots,p'-j'_1-j'_2-2} \\
  {\scriptstyle j =j_1 -j_2 ,j_1 -j_2 +1,\ldots,p -j_1 -j_2 -2},
\end{array}\right\} 
\end{array} \]
The correponding roots of the indicial equations are $s_J:=h_J-h_{J_1}-h_{J_2}$.
A basis for the space of common solutions to $(J_1)$ and $(PJ_1)$ must be contained in 
the set of solutions to $(J_1)$ or $(PJ_1)$
that have leading asymptotics $(z_2-z_1)^{s_J}\log(z_2-z_1)^k$ 
with $J\in \CI_{12}:=\CI_1\cap\CI_2$ for $z_1\rightarrow z_2$. The crucial fact that may be
established by direct calculation is that the roots $s_J$ with $J\in \CI_{12}$ are 
nondegenerate, i.e. there is no $J'\in \CI_1$ such that $s_{J'}=s_J$ and no
$J''\in \CI_2$ such that $s_{J''}=s_J$. It follows that the space of common solutions
of $(J_1)$ and $(PJ_1)$ has dimension less or equal to the cardinality of the set of
$J$ such that the triple $(J,J_2,J_1)$ satisfies the fusion rules. 
Similarly one finds that the
dimension of the space of common solutions to $(J_4)$ and $(PJ_4)$ is bounded from above by the
cardinality of the set of $J$ such that $(J_4,J_3,J)$ obeys the fusion rules.
Consideration of the remaining equations $(J_2)$, $(PJ_2)$, $(J_3)$ and $(PJ_3)$ will not 
lower the bounds on the dimensionalities.\par 
Now one moreover requires the common solutions to $(J_i)$ and $(PJ_i)$, $i=1,2,3,4$ to also obey
(\ref{proj}), i.e. to be of the form (\ref{4ptproj}). These solutions will have asymptotics 
proportional to $(z_2-z_1)^{h_{J_{21}}-h_{J_2}-h_{J_1}}$ for $z_2\rightarrow z_1$ iff 
it has asymptotics $(z_4-z_3)^{h_{J_{21}}-h_{J_4}-h_{J_3}}$ for $z_4\rightarrow z_3$. Therefore only
those $s_J$ which are such that both $(J,J_2,J_1)$ and $(J_4,J_3,J)$ satisfy the
fusion rules can appear as exponents. The dimension of the space of solutions to all 11 equations 
is therefore bounded from above by the dimension of the space of conformal blocks. One thereby
deduces completeness of the set of solutions to all 11 equations that is provided by the 
conformal blocks. \end{proof}
\begin{proposition}\label{reduce}
The theorem is valid iff one has 
\[ \cG^{J_{21}}_{3(21)}(z_1,z_2,z_3,z_4)=
\sum_{J_{23}}F_{J_{21}J_{32}}\sbf{J_3}{J_4}{J_2}{J_1}\cG^{J_{32}}_{(32)1}(z_1,z_2,z_3,z_4)\]
\end{proposition}
{\it Proof:} First of all note that the triple fusion products are defined as formal sums over 
vectors in the homogeneous components $\CH_{J_4}^{(n)}$ of $\CH_{J_4}$. By nondegeneracy of the
form $<,>$ on $\CH_{J_4}^{(n)}$ the theorem is equivalent to the corresponding equation for 
the matrix elements with arbitrary vectors $\xi_4^{(n)}\in\CH_{J_4}^{(n)}$. By repeated application
of the conformal Ward identities it is straightforward to see that any such matrix element
may be expressed as some meromorphic differential operator acting on 
\begin{eqnarray*}
\CF^{J_{21}}_{3(21)}(z_1,z_2,z_3)& := &
<\! v_4,{[}v_3(z_3)\ho{[}v_2(z_2)\ho v_1(z_1){]}_{J_{21}}{]}_{J_4} \! > 
\\
\CF^{J_{32}}_{(32)1}(z_1,z_2,z_3)& := &
<\! v_4,{[}{[} v_3(z_3)\ho v_2(z_2){]}_{J_{32}}\!\ho
v_1(z_1){]}_{J_4} \! >.
\end{eqnarray*}
However to each $\CF^{J_{21}}_{3(21)}$ or $\CF^{J_{32}}_{(32)1}$ one has a 
$\cG^{J_{21}}_{3(21)}$ resp. $\cG^{J_{32}}_{(32)1}$ such that (see (\ref{4ptproj})):\\[1ex]
\hspace*{\fill} $\displaystyle \lim_{z_4\rightarrow\infty}z_4^{2h_4}\cG(z_1,z_2,z_3,z_4) =
\CF(z_1,z_2,z_3).$ \hspace*{\fill}$\Box$\\[1ex]
The proof of Theorem \ref{asso} is thereby completed.\end{proof}
\section{Polynomial equations}
The data $F$ and $\om$ satisfy certain identities. To derive these,
introduce the following conformal blocks to each permutation $(kji)$
of $(321)$ : 
\begin{eqnarray} 
\cG^{J_{ji}}_{k(ji)} &:=& <\! v_{J_4},{[}v_k(z_k)\ho{[}v_j(z_j)\ho
v_i(z_i){]}_{J_{ji}}{]}_{J_4} \!> \\ 
\cG^{J_{kj}}_{(kj)i} &:=& <\! v_{J_4},{[}{[}v_k(z_k)\ho v_j(z_j){]}_{J_{kj}}\ho
v_i(z_i){]}_{J_4} \!>
\end{eqnarray}
These are analytic functions on the universal cover of
$\BA:=\BP^3/{z_i=z_j;i,j=1,2,3}$ of the following form
\begin{eqnarray}
\cG^{J_{ji}}_{k(ji)} &=&
(z_j-z_i)^{\D_{ji}}(z_k-z_i)^{\D_{k(ji)}}H^{J_{ji}}_{k(ji)}
\!\!\left(\frac{z_j-z_i}{z_k-z_i}\right)\\
\cG^{J_{kj}}_{(kj)i} &=&
(z_k-z_j)^{\D_{kj}}(z_j-z_i)^{\D_{(kj)i}}H^{J_{kj}}_{(kj)i}
\!\!\left(\frac{z_k-z_j}{z_j-z_i}\right),
\end{eqnarray}
where $\D_{ji}=h_{ji}-h_j-h_i$, $\D_{k(ji)}=h_4-h_k-h_{ji}$,
$\D_{(kj)i}=h_4-h_{kj}-h_i$ and the functions $H(z)$ are holomorphic and
single-valued in
a neighborhood of $0$.
Consider the region in $\BC^3$ where $(z_2,z_1)$, $(z_3,z_1)$,
$(z_3,z_2)$ are all in $\BC_+^2$.
One then has the following relations between the functions $\cG$:
\begin{eqnarray}
\cG^{J_{21}}_{3(21)}=\om\rb{J_{21}}{J_2\; J_1}\cG^{J_{21}}_{3(12)} & \quad &
\cG^{J_{32}}_{(32)1}=\om\rb{J_{32}}{J_3\; J_2}\cG^{J_{32}}_{(23)1}
\label{brd1}\\ 
\cG^{J_{21}}_{3(21)}=\om\rb{J_4}{J_{21}\; J_3}\cG^{J_{21}}_{(21)3} & \quad &
\cG^{J_{32}}_{(32)1}=\om\rb{J_4}{J_1\; J_{32}}\cG^{J_{32}}_{1(32)} 
\label{brd2} \end{eqnarray}
In addition one has the associativity relations
\begin{eqnarray}
\cG^{J_{ji}}_{k(ji)} &=& \sum_{J_{kj}}F_{J_{ji}J_{kj}}\sbf{J_k}{J_4}{J_j}{J_i}
\cG^{J_{kj}}_{(kj)i} \\
\cG^{J_{kj}}_{(kj)i} &=& \sum_{J_{ji}}
F^{-1}_{J_{kj}J_{ji}}\sbf{J_k}{J_4}{J_j}{J_i}\cG^{J_{ji}}_{k(ji)} 
\end{eqnarray}
Now the expression of $\cG_{3(21)}$ in terms of $\cG_{(21)3}$ may be
  computed in two ways: Either by using (\ref{brd2}) or by a sequence
  of operations that may be symbollically written as  
$3(21)\rightarrow (32)1\rightarrow (23)1\rightarrow 2(31)\rightarrow
2(13) \rightarrow (21)3$. By the linear independence of the
$\cG_{(21)3}^{h_{21}}$ for different $h_{21}$ one gets the following
identity (hexagon):
\[ \om\rb{J_4}{J_3\; J_{21}}=\sum_{J_{32}J_{21}} 
F_{J_{21}J_{32}}\sbf{J_3}{J_4}{J_2}{J_1}\om\rb{J_{32}}{J_3\; J_2}
F^{-1}_{J_{32}J_{31}}\sbf{J_2}{J_4}{J_3}{J_1}\om\rb{J_{31}}{J_3\; J_1}
F_{J_{31}J_{21}}\sbf{J_2}{J_4}{J_1}{J_3}. \]
Similarly one gets  
\[ \om\rb{J_4}{J_{12}\; J_3}=\sum_{J_{21}J_{31}} 
F^{-1}_{J_{32}J_{21}}\sbf{J_3}{J_4}{J_2}{J_1}\om\rb{J_{21}}{J_2\; J_1}
F_{J_{21}J_{31}}\sbf{J_3}{J_4}{J_1}{J_2}\om\rb{J_{31}}{J_3\; J_1}
F^{-1}_{J_{31}J_{32}}\sbf{J_1}{J_4}{J_3}{J_2}. \]
The inverse of $F$ may be calculated in terms of $\om$, $F$ by
representing $(12)3\rightarrow 1(23)$ as the sequence of moves
$(12)3\rightarrow 
(21)3\rightarrow 3(21) \rightarrow (32)1\rightarrow 1(32)\rightarrow
1(23)$. The result is simply 
\begin{equation} F_{J_{21}J_{32}}^{-1}\sbf{J_3}{J_4}{J_2}{J_1}=F_{J_{21}J_{32}}
\sbf{J_1}{J_4}{J_2}{J_3} \label{Finv} \end{equation}
A further important identity may be derived by considering fusion
products of four highest weight states, or equivalently five point conformal blocks:
$[v_4\ho[v_3\ho[v_2\ho v_1]]]$
may be expressed in terms of $[[[v_4\ho v_3]\ho v_2]\ho v_1]$ in two
ways: Either by $4(3(21))\rightarrow (43)(21)\rightarrow ((43)2)1$ or
by $4(3(21))\rightarrow 4((32)1)\rightarrow (4(32))1\rightarrow
((43)2)1$. This leads to the identity (pentagon)
\begin{equation}
F_{J_{321}J_{43}}\sbf{J_4}{h}{J_3}{J_{21}}
F_{J_{21}J_{432}}\sbf{J_{43}}{J_5}{J_2}{J_1} = 
\sum_{J_{32}} 
F_{J_{21}J_{32}}\sbf{J_3}{J_{321}}{J_2}{J_1}
F_{J_{321}J_{432}}\sbf{J_4}{J_4}{J_{32}}{J_1}
F_{J_{32}J_{43}}\sbf{J_4}{J_{432}}{J_3}{J_2}.
\label{penta} \end{equation}
If one then considers
eqn. (\ref{penta}) in the special cases $J_1=(0,1/2)$,
$J_{21}=(j_2',j_2^{}+1/2)$ and $J_1=(1/2,0)$, $J_{21}=(j_2'+1/2,j_2^{})$, 
one finds that it allows to express the F-matrices with
$J_1=(j_1',j_1^{}+1/2)$ (or $J_1=(j_1'+1/2,j_1^{})$) in terms of those
with $J_1=(i_1',i_1^{})$, $i'_1\leq
j'_1$, $i_1\leq j_1$. Equation (\ref{penta}) therefore 
\[ \mbox{uniquely determines}\quad 
F_{J_{21}J_{32}}\sbf{J_3}{J_4}{J_2}{J_1}\quad\mbox{in terms of}\quad 
F_{J_{21}J_{32}}\sbf{J_3}{J_4}{J_2}{(0,1/2)},\,\,
F_{J_{21}J_{32}}\sbf{J_3}{J_4}{J_2}{(1/2,0)}.\]
\section{Conformal bootstrap}
Now one is in the position to apply the results of \cite{CGR} to determine the matrices $F$ 
explicitely. The result is
\begin{thm} There is a number $g\rb{J_2}{J_3\; J_1}$ for any triple $(J_3,J_2,J_1)$ that
satisfies the fusion rules such that
\begin{equation} F_{J_{21}J_{32}}\sbf{J_3}{J_4}{J_2}{J_1}=
\frac{g\rb{J_3}{J_{32}\;J_2}g\rb{J_{32}}{J_4\;J_1}}{ 
g\rb{J_2}{J_{21}\; J_1}g\rb{J_3}{J_4\; J_{21}}}
\gsj{J_1}{J_3}{J_2}{J_4}{J_{21}}{J_{32}},  \label{Fex}\end{equation}
where $\gsj{J_1}{J_3}{J_2}{J_4}{J_{21}}{J_{32}}$ is defined as
\[ \gsj{J_1}{J_3}{J_2}{J_4}{J_{21}}{J_{32}}:= 
(-)^{2j'_2(j_{12}+j_{23}-j_2-j_4)+2j_2(j'_{12}+j'_{23}-j'_2-j'_4)}
\sj{j^{}_1}{j^{}_3}{j^{}_2}{j^{}_4}{j^{}_{21}}{j^{}_{32}}
\sjm{j'_1}{j'_3}{j'_2}{j'_4}{j'_{21}}{j'_{32}} \]
in terms of the restricted q-6j symbols defined in \cite{KR}.
\end{thm}
In order to see that the results of \cite{CGR} may be applied here, I
will briefly review the overall strategy:\par
In the case of $J_1=(0,1/2)$ or $J_1=(1/2,0)$ one gets second order differential equations from
null vector decoupling, which may be reduced to the hypergeometric differential equation.
The matrix $F$ for this case is thereby found in terms of Gamma-functions. It is then shown in
\cite{CGR} how to determine $g\rb{J_2}{J_3\; J_1}$ such that for irrational $c$ and the special 
cases $J_1=(0,1/2)$ or $J_1=(1/2,0)$ equation (\ref{Fex}) holds (of course with unrestricted
q-6j).\par
However, one may explicitely check that in these cases the elements of $F$ matrices and
q-6j symbols do not vanish in the limit $c\rightarrow c_{p'p}$, provided the triples 
$(J_1,J_2,J_{12})$, $(J_{12},J_3,J_4)$, $(J_2,J_3,J_{23})$, $(J_{23},J_1,J_4)$ satisfy the
fusion rules. Equation (\ref{Fex}) will therefore hold for $c=c_{p'p}$ 
if the q-6j symbols are simply taken to be the restricted q-6j as defined in \cite{KR}. 
Validity of equation (\ref{Fex}) for general $J_1$ will then follow recursively from the pentagon
equation.
\section{Appendix}
\subsection{Proof of Lemma \ref{fuchsian}}
Call differential operators $\CD$ fuchsian iff $\CD f(z)=0$ is of fuchsian type.\par
It suffices to show that 
\begin{equation}\label{fuchs}
\CL_{-n_k}\ldots\CL_{-n_1}\cG=\CD_{-\pi}\cG,\quad\mbox{where}\quad
\CD_{-\pi}\quad\mbox{is fuchsian}.\end{equation}
 For notational simplicity consider the equations from the
nullvectors on $\CV_{J_1}$ and let $z:= z_1$, $\pa:=\pa_1$. 
(\ref{fuchs}) may be proved by induction on $k$: 
It is easy to see that it holds for $k=1$ by direct calculation using (\ref{4ptproj}): 
One has 
\begin{equation}\label{lex} \CL_{-m}\cG=\CD_{-m}\cG:= \left( \frac{c_2}{z_{21}^{m}}
+\frac{1}{z_{21}^{m-1}}\frac{z_{41}z_{31}}{z_{42}z_{23}}\pa
+(2\leftrightarrow 3)+(2\leftrightarrow 4)
\right) \cG, \end{equation} with some constants $c_2,c_3,c_4$,
which is easily checked to be fuchsian. An important property of $\CD_{-m}$ is that its 
coefficients are functions of $z_{21},z_{31},z_{41}$ only.\par 
Now assume that $\CD_{-\pi}$ is in the set $\CF^{(l)}$ of differential operators of the form
\begin{equation}\label{form} \CD_{-\pi}=\sum_{k=0}^l q^{(k)}\pa^k,\qquad 
q^{(k)}=\frac{p^{(k)}(z_{21},z_{31},z_{41})}{(z_{21}z_{31}z_{41})^{n-k}},  \end{equation}
where $p^{(k)}(z_{21},z_{31},z_{41})$ is polynomial of order $2(l-k)$ in $z_1$, and consider
$\CL_{-m}\CD_{-\pi}\cG={[}\CL_{-m},\CD_{-\pi}{]}\cG+\CD_{-\pi}\CD_{-m}\cG$.
The differential operator 
$\CD_{-\pi}\CD_{-m}$ is easily seen to be in $\CF^{(l+m)}$, so it remains to consider the first
term. Write $\CL_{-m}=r_m+s^i_{m-1}\pa_i$, where $i$ is summed over $i=2,3,4$. 
\[ {[}\CL_{-m},\CD_{-\pi}{]}=\sum_{k=0}^{l}(\CL_{-m}q^{(k)})\pa^k+(\CD_{-\pi}s^i_{m-1})\pa_i
+\CD_{-\pi}r_{m}. \]
The last term is trivially fuchsian, for the second one note that
$q^{(k)}(\pa^k s_{m-1}^i)\pa_i$ is proportional to $q^{(k)}s_{m+k-1}^i\pa_i$, which is again 
fuchsian when acting on $\cG$, see (\ref{lex}). It remains to show that  
\[ \CL_{-m}q^{(k)}:= r_m q^{(k)}+s_{m-1}^i\pa_i q^{(k)}=
\frac{r^{(n+m-k)}(z_{21},z_{31},z_{41})}{(z_{21}z_{31}z_{41})^{n+m-k}}, \]
where $r^{(n+m-k)}(z_{21},z_{31},z_{41})$ is of order $2(l+m-k)$ in $z_1$. This reduces be the 
verification that  $((z_{41}z_{31})^{m-1}\pa_2+(z_{41}z_{21})^{m-1}\pa_3+(z_{21}z_{31})^{m-1}\pa_4)
p^{(k)}(z_{21},z_{31},z_{41})$ is (a) of order $2(l+m-k)$ and (b) annihilated by 
$\sum_{i=1}^4\pa_i$. For (a) note that because of $\sum_{i=1}^4\pa_i p^{(k)}=0$, 
$(\pa_2+\pa_3+\pa_4)p^{(k)}$ is of order $2(l-k)-1$ in $z_1$. (b) follows from the fact that 
$\sum_{i=1}^4\pa_i$ commutes with $\CL_{-m}$.$\Box$
\subsection{Proof of Lemma \ref{indicial}}
It suffices to show that the indicial equation for the differential operator 
$\CD_{-\pi}$ is given by
\[ \prod_{i=1}^{k}\left(\D-\sum_{j=1}^{i-1}n_j+h_2(1-n_i)\right)=0. \]
Use induction on $k$: From the previous lemma one has 
\[ \CD_{-\pi}=\sum_{k=0}^l q^{(k)}\pa_i^{(k)} \qquad\mbox{with}\quad 
q^{(l-k)}=z_{ij}^{-(l-k)}a^{(k)}+{\cal O}(z_{ij}^{-(l-k)+1}) \]
such that the indicial equation is $\sum_{k=0}^l a^{(k)}(s)_k=0$,
where $(s)_k:= s(s-1)\ldots(s-k+1)$. Now write
\[ \CL_{-m}\CD_{-\pi}\cG=\sum_{k=0}^l\left( (\CL_{-m}q^{(k)})\pa_i^k+q^{(k)}(s_{m-1}^j\pa_i^k\pa_j)
\right)\cG \]
The first term contributes $(k+h_j(m-1))a^{(k)}$. To evaluate the second, use (\ref{4ptproj})
and observe that because of ($\tilde{z}:= \frac{z_{21}z_{43}}{z_{31}z_{42}}$)
\[ 
\pa_2 F(\tilde{z})=\frac{z_{41}z_{31}}{z_{42}z_{23}}\pa_1F(\tilde{z})\quad  
\pa_3 F(\tilde{z})=\frac{z_{41}z_{21}}{z_{43}z_{32}}\pa_1F(\tilde{z})\quad  
\pa_4 F(\tilde{z})=\frac{z_{21}z_{31}}{z_{24}z_{43}}\pa_1F(\tilde{z}),
\]
only the $j=2$ term is relevant,
so that the contribution to the indicial equation is $-a^{(k)}(s)_{k+1}$.
Collecting terms one therefore finds the indicial equation to be 
\[ 0=\sum_{k=0}^{l}a^{(k)}((s)_k(k+h_j(m-1))(s)_k-a^{(k)}(s)_{k+1}=
(h_j(m-1)-s+l)\sum_{k=0}^{l}a^{(k)}(s)_k.\]
The claim follows.
\newcommand{\CMP}[3]{{\it Comm. Math. Phys. }{\bf #1} (#2) #3}
\newcommand{\NP}[3]{{\it Nucl. Phys. }{\bf B#1} (#2) #3}
\newcommand{\PL}[3]{{\it Phys. Lett. }{\bf B#1} (#2) #3}
\newcommand{\PRL}[3]{{\it Phys. Rev. Lett. }{\bf #1} (#2) #3}
\newcommand{\AP}[3]{{\it Ann. Phys. (N.Y.) }{\bf #1} (#2) #3}
\newcommand{\LMJ}[3]{{\it Leningrad Math. J. }{\bf #1} (#2) #3}
\newcommand{\FAA}[3]{{\it Funct. Anal. Appl. }{\bf #1} (#2) #3}
\newcommand{\PTPS}[3]{{\it Progr. Theor. Phys. Suppl. }{\bf #1} (#2) #3}
\newcommand{\LMN}[3]{{\it Lecture Notes in Mathematics }{\bf #1} (#2) #2}

\end{document}